\documentclass[aps,prb,preprint,superscriptaddress,usenames,dvipsnames]{revtex4-1} 

\usepackage{graphicx}

\usepackage{siunitx}
\usepackage{textcomp}
\usepackage{todonotes}
\usepackage{color}
\usepackage[normalem]{ulem}
\usepackage[hidelinks]{hyperref}

\newcommand{\mos}{MoS\ensuremath{_2}}
\newcommand{\mose}{MoSe\ensuremath{_2}}
\newcommand{\mote}{MoTe\ensuremath{_2}}

\newcommand{\jref}[1]{Ref.\ \cite{#1}}
\newcommand{\jfig}[1]{Fig.\ \ref{fig:#1}}

\newcommand{\jexp}[2]{\ensuremath{#1 \times 10^{#2}}}
\newcommand{\ee}{\ensuremath{E_0}}
\newcommand{\kp}{\ensuremath{k_{\parallel}}}
\newcommand{\KP}{\ensuremath{{\mathbf k}_{\parallel}}}   

\begin{document}


\title{
    Unoccupied bands in the molybdenum dichalcogenides MoS$_2$, MoSe$_2$, and MoTe$_2$
}

\author{J.\ Jobst}\affiliation{Huygens-Kamerlingh Onnes Laboratorium, Leiden Institute of Physics, Leiden University, Niels Bohrweg 2, P.O. Box 9504, NL-2300 RA Leiden, The Netherlands}
\affiliation{Department of Physics, Columbia University, New York, NY, US}

\author{E.E.\ Krasovskii}
\affiliation{Universidad del Pais Vasco/Euskal Herriko Unibertsitatea, 20080 Donostia/San Sebastián, Basque Country, Spain}
\affiliation{IKERBASQUE, Basque Foundation for Science, E-48013 Bilbao, Spain}
\affiliation{Donostia International Physics Center (DIPC), E-20018 San Sebasti\'an, Spain}

\author{R.\ Ribeiro}
\affiliation{Department of Physics, Columbia University, New York, NY, US}

\author{T.A.\ de Jong}
\affiliation{Huygens-Kamerlingh Onnes Laboratorium, Leiden Institute of Physics, Leiden University, Niels Bohrweg 2, P.O. Box 9504, NL-2300 RA Leiden, The Netherlands}

\author{C.R.\ Dean}
\affiliation{Department of Physics, Columbia University, New York, NY, US}

\author{R.M.\ Tromp}
\affiliation{IBM T.J.Watson Research Center, 1101 Kitchawan Road, P.O.\ Box 218, Yorktown Heights, New York 10598, USA}
\affiliation{Huygens-Kamerlingh Onnes Laboratorium, Leiden Institute of Physics, Leiden University, Niels Bohrweg 2, P.O. Box 9504, NL-2300 RA Leiden, The Netherlands}

\author{S.J.\ van der Molen}
\email{molen@physics.leidenuniv.nl}
\affiliation{Huygens-Kamerlingh Onnes Laboratorium, Leiden Institute of Physics, Leiden University, Niels Bohrweg 2, P.O. Box 9504, NL-2300 RA Leiden, The Netherlands}

\date{\today}

\begin{abstract}
We present angle-resolved reflected electron spectroscopy (ARRES) data for the three molybdenum-based transition metal dichalcogenides (TMDs) \mos, \mose, and \mote. To follow the changes as the series moves from S to Se to Te  in more detail, we determine accurate IV-spectra for monolayers and bulk TMDs. These experimental data sets are then compared  with theoretical predictions for both the unoccupied band structure and the scattering density of states. We find good agreement, especially for lower energies where inelastic effects are relatively unimportant. Furthermore, we identify a series of interlayer resonances for which the dependence of the hybridization effects on the layer
count is observed. Although these resonances bear similarity to interlayer resonances in hBN and graphene, they differ in their character, being dominated by unoccupied $d$-states of the chalcogen-atoms. The unoccupied states studied and analyzed here play a key role in all processes that require an electron to temporarily reside in a state above the vacuum level, such as in photoemission and secondary electron emission experiments. 

\end{abstract}


\maketitle

Van der Waals (vdW) systems, materials consisting of weakly interacting layers, exhibit a wealth of physics phenomena and a rich perspective for applications~\cite{Geim2013, Wu2021, Zollner2025, Cao2018, Yankowitz2019}. Specifically, they hold the promise that new materials with novel - and even - pre-programmed properties may be designed and constructed by combining single layers from different parent materials in a specific configuration. In fact, there is already an impressive richness for stacked layers of the same material. Examples include the transition from linear to quadratic dispersion upon going from monolayer to bilayer graphene \cite{novoselov-QHE}, the appearance of superconductivity for two graphene layers stacked at a `magic' twist angle \cite{herrero2018-superconductivity}, and the shift from direct to indirect semiconductor behavior vs. thickness, for several transition metal dichalcogenides (TMD) \cite{Mak2010}.
By far the \lq simplest\rq\ vdW building blocks are formed by unit layers that have all atoms and bonds within the same plane, such as graphene and single-layer hBN. In a unit vdW layer of a TMD, the transition metal atoms are sandwiched between two planes of chalcogen atoms (S, Se or Te). This results in more complex electronic and crystallographic properties and a relatively large distance between the planes of the transition metal atoms. Here, we study the series of TMD systems \mos, \mose, and \mote, of different thicknesses, all in the 2H phase, using angle-resolved reflected electron spectroscopy (ARRES). This technique allows one to measure the dispersion relations of unoccupied electron states for spatial structures as small as $10\times 10\,\text{nm}^2$~\cite{Jobst-ARRES, Jobst-ARRES-GonBN}. The latter property makes ARRES an ideal tool to characterize the small, locally overlapping flakes that are typically used in vdW materials research. The unoccupied bands probed in ARRES are relatively high in energy, i.e., above the vacuum level, and play an important role in secondary electron emission and photo-emission. For example, they can be related to the so-called final states in angle-resolved photo-emission spectroscopy (ARPES)~\cite{Strocov2001-photoemission-graphite, Tebyani2023}. In the one-step theory of photoemission the wave function describing the angular distribution of the photoemission intensity is the time-reversed LEED state, which is the scattering wave function in the LEED experiment. Material properties such as crystal structure and chemical composition  determine the Hamiltonian of the system. The LEED wave functions are scattering eigenfunctions of this Hamiltonian, which establishes a firm link between the structure of the material and the energy-angular distribution of the reflected intensity \cite{Adawi1964, Mahan1970, Pendry1976, Vilkov1999}.
In ARRES, the local electron reflectivity is recorded as a function of both electron energy and in-plane momentum. The technique is closely related to the $T(E)$-VLEED method introduced by Strocov \emph{et al}.~\cite{Strocov1997, Strocov2000-3D-unoccupied}. ARRES has been benchmarked for multilayer graphene and hBN, where a set of unoccupied resonances are observed that stem from coupling between interlayer resonances~\cite{Hibino2008-full-IVs,hibino-graphene-hBN,feenstra-interlayer, PatakPRB2025}. Consequently, the number of resonances seen is directly related to the layer count: for $n$ layers, one will find $n - 1$ minima in the ARRES spectrum at the $\Gamma$-point, the so-called `LEEM IV-spectrum' (reflectivity as a function of electron energy). The energy splitting between minima is a direct measure of the interaction between individual interlayer resonances. In the limit of an infinite crystal the transmission resonances merge to form unoccupied conducting bands \cite{Krasovskii2025-negative-transit}. Note that the localized resonant states do not influence material properties in the ground state because they are unoccupied. They are populated by the incoming probing electrons, but the intensity of the probing beam is too low to cause tangible changes in the material. The same is true for traditional ARPES: the photocurrent can be considered a negligible perturbation, while significant band bending (surface photovoltage) can be caused due to absorption of the light in ARPES. For the more complex TMD family, however, such a comprehensive, let alone intuitive understanding of the unoccupied band structure is still largely absent~\cite{Neu2023}.

In this contribution, we present low-energy electron microscopy (LEEM) data from the three molybdenum-based TMDs \mos, \mose, and \mote, obtained by mechanical exfoliation. We compare ARRES spectra for monolayers and bulk with theoretical predictions for both the unoccupied band structure and the scattering probability and find good agreement, especially for the lower energies. We use this comparison to demonstrate that much of the complex behavior of electron reflectivity in TMDs can indeed be understood in a framework of unoccupied bands, propagating states and scattering probabilities. Furthermore, we are able to pinpoint unoccupied interlayer states in these TMDs. We discuss their orbital character as well as the role of hybridization effects as the number of layers increases.

\begin{figure*}[ht]
	\centering  
	\includegraphics[width=\textwidth]{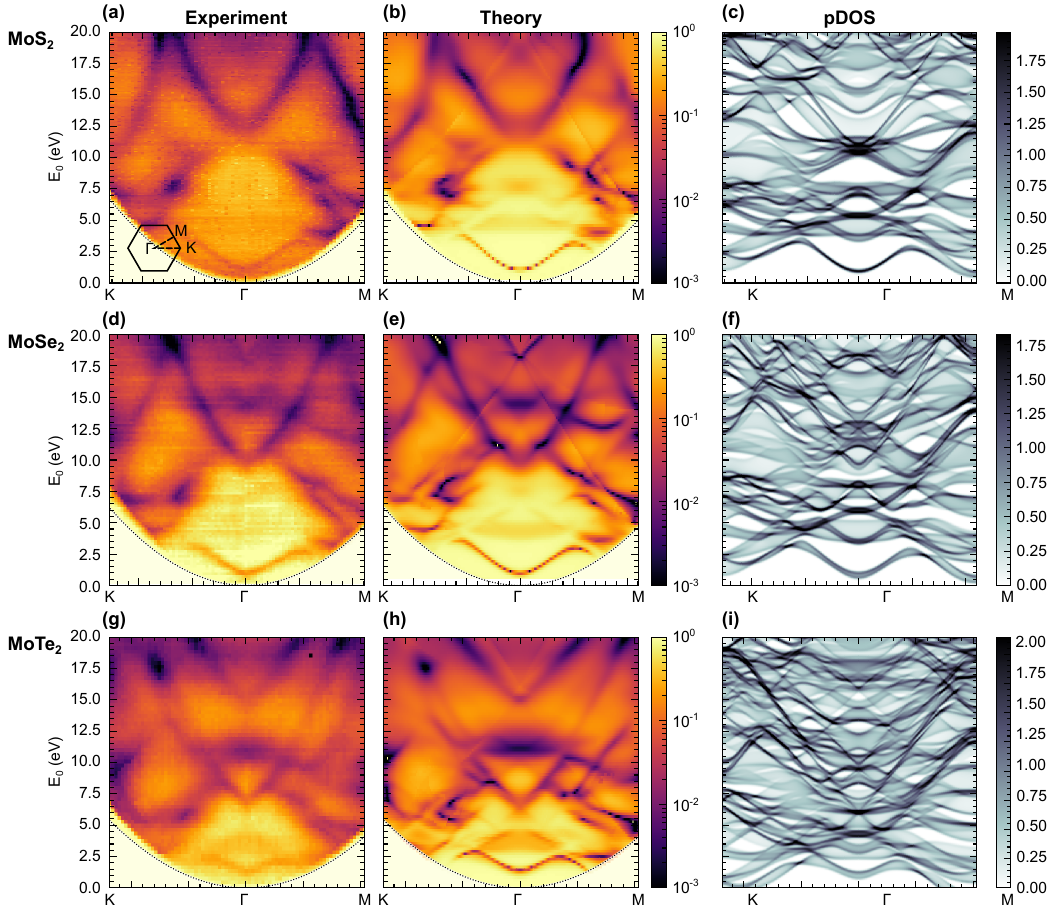}
	\caption{
		(a) ARRES measurement on bulk \mos\ along the high-symmetry directions K-$\Gamma$-M. High intensity (bright colors) correspond to high electron reflectivity, dark colors to low reflectivity. The dashed line indicates the electron vacuum dispersion below which no ARRES data can be acquired.
		(b) Electron intensity reflected from a semi-infinite \mos\ slab as calculated using the augmented plane wave method (see main text). 
		(c) Calculated \KP-projected density of states (pDOS) above the vacuum level. White areas correspond to band gaps.  
		(d-e-f) The same series for bulk \mose.		
		(g-h-i) The same series for bulk \mote.}
		\label{fig:bulk} 
\end{figure*}

\section{Experimental}
All LEEM measurements are performed using the aberration correcting \mbox{ESCHER} LEEM facility \cite{schramm2011low} which is based on a commercial SPECS P90AC instrument~\cite{Tromp-AC1, Tromp-AC2}. The set-up does not only provide high spatial resolution, but also a high energy resolution of $\approx 0.25$ eV FWHM due to its cold-field-emission electron gun.
To obtain the TMD samples, thin few-layer crystals were mechanically exfoliated from bulk crystals of \mos, \mose, and \mote\ (all 2H phase) using Scotch tape, and deposited onto silicon chips with a thermal oxide of ~300\,nm thickness as described in \jref{novoselov-QHE}. Promising flakes with monolayer areas were identified utilizing optical interference contrast and transferred onto a conductive silicon chip, using a polycarbonate (PC) coated PDMS stamp as described in \jref{Zomer2014,Wang2013}. Those chips were cleaned for 30\,s in oxygen plasma prior to the deposition of the flakes. This treatment ensures that the chips are clean and enhances the adhesion of the TMD flakes while not increasing the thickness of the $\sim$1\,nm of native oxide. Note that a thicker oxide would cause charging in LEEM experiments and thus is to be avoided. Finally, the PC residues are washed off in ultra-pure chloroform and quickly blown dry with a nitrogen gun. 
The samples are then transferred into the LEEM instrument (base pressure of \jexp{5}{-10}\,mbar) and outgassed over night at 400\textcelsius\ to remove water and adsorbed hydrocarbons. Suitable bulk flakes have first been identified using the PEEM-mode (photo-electron emission microscopy) of the LEEM instrument before further investigation with LEEM-mode. Suitable flakes are either flakes that have been selected from optical imaging contrast (see above) and were identified in PEEM by their shape or flakes that display PEEM intensity variations and fold lines at their edges that indicate areas of few-layer material. All microscopy has been performed below a pressure of $2\times10^{-9}$\,mbar and at \SI{450}{\celsius} to prevent the formation of hydrocarbon-based contaminants under the electron beam. We do not observe any degradation or phase changes of the flakes at this annealing temperature over the course of days. Phase changes are not expected since the  2H crystals represent the thermodynamically most stable form of the TMDs. Compared to XPS or similar methods, LEEM-based spectroscopy is less sensitive to small concentrations of atomic vacancies, which might form at these temperatures, and we cannot fully rule them out despite the ultra-high vacuum conditions. The agreement with simulations assuming perfect crystals (see below) further indicates that annealing at these conditions does not noticeably change the TMDs. All LEEM images have been corrected for detector-induced artifacts by subtracting a dark count image and dividing by a gain image before further analysis as described in \jref{DeJong2019}. All images from which reflectivity spectra are extracted are integrated for 250\,ms, the insets in \jfig{IVs} are integrated for 4\,s. We note that, throughout this article, the energy \ee\ is the electron energy in vacuum, and the relevant electronic states are scattering states.

\section{Results}
Figure \ref{fig:bulk} compares ARRES measurements performed on bulk \mos, \mose, and \mote\ with theoretical calculations. The data in \jfig{bulk}(a), \jfig{bulk}(d) and \jfig{bulk}(g) have been recorded in \lq diffraction mode\rq, which means by taking a series of LEED patterns of the area of choice while sampling energy \ee\ and in-plane momentum \kp\ of the incoming electrons. The intensities of the (0,0) spots are extracted from the LEED images and are plotted as the ARRES spectrum $I(\ee,\kp)$ in Fig.\ \ref{fig:bulk} as in Refs.\ \cite{Strocov1997, Jobst-ARRES}. Note that electrons with energy $E_0$ below  the free electron dispersion $(\hbar\cdot k_\parallel)^2/2m_0$ (dashed parabolas in the figure), where $m_0$ is the free electron mass, cannot reach the sample in our experiment, due to insufficient kinetic energy along the surface normal. Their reflection probability is therefore unity (a situation called `mirror mode reflection') \cite{Jobst-ARRES}. Note that selection of the (0,0) LEED beam in the diffraction plane, which is also an energy-dispersive plane in the LEEM instrument, automatically removes inelastically scattered electrons from the recorded LEED intensities.

In a simplified picture, working remarkably well for multilayer graphene and hBN, minima in ARRES spectra  correspond to a high density of states above $E_{\rm vac}$, enhancing the transmission probability of electrons into the material~\cite{Wicki2016-mapping, Geelen2019, Neu2021}. Equivalently, reflection maxima indicate band gaps (final-state gaps in ARPES terminology~\cite{Miller2015-resolving}), which increase the probability for an electron to be reflected. Hence, the ARRES spectra are expected to be a fingerprint of the \KP-projected density of states (pDOS) of a material. In general, however, a full calculation of the problem should include the scattering probabilities for each propagating state. 

The features in the ARRES spectra in \jfig{bulk}(a,d,g) are determined by the interaction of the incoming electron plane wave with the potential landscape arising from the atoms in the TMD layers. The simulations that we compare to our experiments have to solve this quantum mechanical scattering problem for the TMD structures considered. They were performed with a full-potential linear augmented plane wave method based on a self-consistent crystal potential obtained within the local density approximation, as explained in Ref.\ \onlinecite{krasovskii1999-augmented}. The \emph{ab-initio} reflectivity spectra are obtained with the all-electron Bloch-wave-based scattering method described in Ref.\ \onlinecite{krasovskii2004-augmented}. The extension of this method to stand-alone two-dimensional films of finite thickness was introduced in Ref.\ \onlinecite{nazarov2013-scattering-2D}. Here, it is straightforwardly applied to the case of finite incidence angle to represent the experimental geometry. For few-layer TMDs we use the same interlayer distances as known for bulk crystals since no change is expected due to the weak van-der-Waals bonding. In our calculations, an absorbing optical potential was introduced to account for inelastic scattering. For this we used the same energy-dependent optical potential $V_\mathrm{i}(E)$ for all materials. It has previously been calculated for the similar TMD WSe$_2$~\cite{Silkin2018}. In addition, a Gaussian broadening of \SI{1}{\electronvolt} is applied to account for the combined broadening effects in the experiment. 

The results of these calculations are shown in Figs.~\ref{fig:bulk}(b), \ref{fig:bulk}(e), and \ref{fig:bulk}(h) for \mos, \mose, and \mote, respectively. The computed results agree very well with the measured ARRES data in Figs.~\ref{fig:bulk}(a), \ref{fig:bulk}(d), and \ref{fig:bulk}(g), over the whole \ee\ and \kp\ range. To achieve the agreement, the origin of the theoretical kinetic energy scale was adjusted to match the experiment and the energy scale was stretched by $\approx5\%$ to compensate for the underestimated energies of unoccupied states. This underestimate is a consequence of using the local density approximation instead of the true quasi-particle self-energy. These corrections are intrinsic to the simulations and are of the expected magnitude but are testament to the crucial importance of experimental data as a benchmark for theory.

The excellent correspondence of experimental and theoretical ARRES spectra enables us to investigate whether a direct relationship between the spectra and the (calculated) unoccupied pDOS exists in these TMDs. The calculated curves for Bloch waves propagating into the TMD are shown in Figs.~\ref{fig:bulk}(c), \ref{fig:bulk}(f), and \ref{fig:bulk}(i) for \mos, \mose, and \mote, respectively. Especially for the lower energies, a convincing correspondence is observed. Specifically, the dispersive lowest band shows up in the ARRES and pDOS plots for all TMDs considered. For higher energies, a detailed comparison between experiment and theory becomes increasingly challenging since the large number of propagating Bloch states, often overlapping in their pDOS, hinders a precise identification. For the more complex TMDs this effect is more pronounced than for simpler VdW systems such as graphite and hBN~\cite{Strocov2000-3D-unoccupied, Jobst-ARRES, Jobst-ARRES-GonBN}. This trend is also visible within the Mo-based TMD series investigated here, where it is most challenging for \mote\ with the heaviest atom (Te) in the series.\\

\begin{figure*}[t]
	\centering  
	\includegraphics[width=\textwidth]{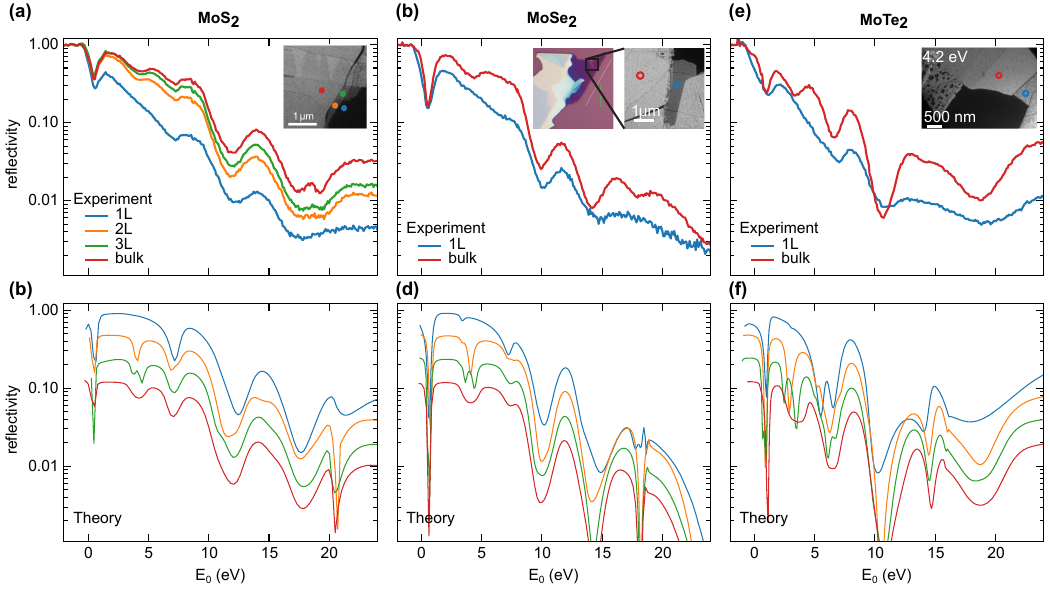}
	\caption{
		(a,b,c) Reflectivity spectra (IV-curves) recorded from bright-field LEEM images of \mos, \mose, and \mote, respectively. The monolayer spectra lack the minimum between 2\,eV and 6\,eV that spectra from thicker areas exhibit. The regions where the spectra were recorded are indicated in the inset LEEM images.
		(d,e,f) Simulated reflectivity spectra for \mos, \mose, and \mote, respectively, describe the experimental data well over the full energy range. In particular, they exhibit the same missing minimum for all monolayer TMDs. The curves for monolayer, bilayer, trilayer, and bulk have colors corresponding to (a-c) and thicker layers are vertically shifted downward for clarity in the theoretical plots.}
		\label{fig:IVs}
\end{figure*}

As mentioned, for multilayer graphene and hBN the lowest-energy ARRES region (0 to 5~eV) contains information on the exact layer count and on the coupling strength between neighboring (interlayer) resonances~\cite{Hibino2008-full-IVs, feenstra-interlayer, Jobst-ARRES, Jobst-ARRES-GonBN}. The question arises whether this simple picture still holds for TMDs, with their more complex unit cells. 

Figures~\ref{fig:IVs}(a), \ref{fig:IVs}(b), and \ref{fig:IVs}(c) show reflectivity spectra (`LEEM IV's') for different layer counts of \mos, \mose, and \mote, respectively. These data are acquired from the areas shown in the insets. Note that the bulk spectra in \jfig{IVs} correspond to vertical cuts at the $\Gamma$-point in Figs.~\ref{fig:bulk}(a), \ref{fig:bulk}(d), and \ref{fig:bulk}(g). 
All curves for \mos\ appear rather similar at first sight. Still, three things stand out~\cite{Neu2023, DeJong2018pssb}.  First, the overall reflectivity 
increases with thickness, most likely due to improving material flatness, rigidity and/or crystallinity as the number of layers increases \cite{Neu2025}. Second, the spectra for bilayer and thicker \mos\ are nearly identical apart from a minimum developing around 21~eV for \mos\ (18 eV and 15 eV for \mose\ and \mote, respectively). These minima appear when the electrons have sufficient energy to scatter into the off-normal, first-order LEED spots and thus reduce the intensity at the $\Gamma$ point. This effect becomes more pronounced as the layer number increases due to increased flatness. Finally, the monolayer curve is markedly different from the rest; it is the only one not showing a minimum at $\approx 4$\,eV. Similar behavior is found for \mose\ [\jfig{IVs}(b)] and \mote\ [\jfig{IVs}(c)], where we compare the monolayer to the bulk case. Remarkably, we can understand the minima around 4\,eV in terms of interlayer resonances, just as for graphene and hBN. For this, we refer to the reflectivity calculations shown in Figs.~\ref{fig:IVs}(b), \ref{fig:IVs}(d), and \ref{fig:IVs}(f) for \mos, \mose, and \mote, respectively (note that these curves have been offset for clarity in contrast to the experimental data). Focusing on \mos\ again, we do not find any feature around $4$ eV for the monolayer, while a minimum does occur for the double layer. For three layers, this minimum splits into two, and further evolves into a broad minimum for bulk \mos~\cite{DeJong2018pssb}. These results, which are also found for \mose\ and \mote, are in line with what is seen for graphene and hBN and are indeed related to interlayer resonances in these (more complicated) TMD systems, as discussed next.

\begin{figure}[t]
	\centering  
	\includegraphics[width=\columnwidth]{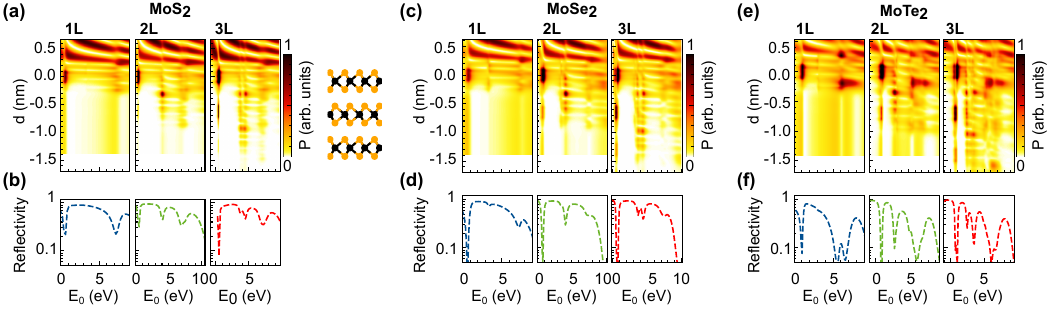}
	\caption{(a) Electron density $P$ inside and outside the \mos -system for monolayer, bilayer and trilayer. It is calculated for a series of incoming electron waves with landing energies $0 < E_0 < 10$\,eV. The first Mo-plane is set at $d=0$. The electron waves are coming from the top and their reflection leads to an interference pattern in front of the material ($d > 0.3$ nm), which depends on energy (wavelength). Inside the material, resonant states with higher density (in red) are seen. A schematic of the triple-layer \mos\ structure with Mo-\ and S-planes aligned with the correct depth values $d$ is show for comparison. (b) Calculated reflectivity spectra for monolayer, bilayer and trilayer \mos\  [same as in Fig. \ref{fig:IVs}(b)]. Panels (c) and (d) shows the same quantities for \mose, and (e) and (f) for \mote.}
	\label{fig:density} 
\end{figure}

Figure~\ref{fig:density}(a) shows calculated probability density profiles $P$ in the scattering states of \mos\ \emph{within} the crystal as well as \emph{in front of it} in the energy interval 0 to 10\,eV, for (from left to right) mono-, bi-, and trilayer \mos. 
The calculated (dashed) reflectivity curves are shown in the three panels of Figure \ref{fig:density} (b) for \mos\ [same curves as in Fig. \ref{fig:IVs}(b)].

Let us first inspect the reflectivity curves, where a sharp minimum is seen around $E_0 = 0.5$\,eV for all thicknesses considered. This feature corresponds to the lowermost band at the $\Gamma$ point for the semi-infinite  \mos, see Figs.~\ref{fig:bulk}(a)--\ref{fig:bulk}(c)]. For this energy, the density plot for the monolayer exhibits a high electron density region \emph{within} the \mos\ layer (namely around the Mo-plane). Around the same energy, similar density features are found for the multilayers, with additional electron weight at the subsequent Mo-plane(s). In other words, a resonant state that is located around the consecutive Mo planes is responsible for the minimum in reflectivity at $E_0 \approx 0.5$~eV. The position of this minimum does not change much when going from one to three layers or to bulk material.\\ 
The opposite holds for the resonance at $E_0 \approx 4$ eV, which does not occur for the monolayer, suggesting that it is an interlayer resonance. Indeed, for the bilayer calculations in Fig.~\ref{fig:density}(a), we find that most of the density of this resonant state is located between the two \mos\ layers. (Note that no interlayer state is formed between the first \mos\ layer and the oxidized silicon substrate due to the latter's amorphous nature.) For the trilayer, the two original interlayer resonances are coupled, leading to a splitting of the minimum in the calculated reflectivity curve in Fig.~\ref{fig:density}(b). This splitting is also visible in the density plot in Fig.~\ref{fig:density}(a) around 4\,eV, which also highlights the finite electron densities located between the three respective TMD layers. The results for \mose\ shown in Fig.~\ref{fig:density}(c) and (d) as well as for \mote\ in Fig.~\ref{fig:density}(e) and (f) display qualitatively similar behavior. We note that such general features are similar to what is seen in multilayer hBN and graphene around a landing energy $E_0 \approx 3$\,eV. Hence, also these TMD interlayer resonances make it, in principle, possible to count TMD layers by LEEM. The coupling energy between the TMD layers, however, is significantly smaller than for hBN and multilayer graphene, as the interlayer distance is much larger than in those materials.  Unfortunately, this also makes it harder to measure the splitting in TMDs: the experimental curve for three layers of \mos\ in Fig.~\ref{fig:IVs}(a) is too smeared to distinguish two minima convincingly. Spectroscopy using LEEM with improved energy resolution is required to resolve this issue~\cite{Tromp2023}.\\
Finally, we find features around $7$\,eV that correspond to resonant states that are delocalized over the entire \mos , \mose\ or \mote\ thickness. The related minimum in reflectivity is rather independent of layer count.\\
We note that in general the calculated probability densities for \mose\ and \mote\ are similar to \mos\ for the energies considered, although the additional states for especially \mote\ do influence the detailed picture.\\

\begin{figure*}[t]
	\centering  
	\includegraphics[width=\textwidth]{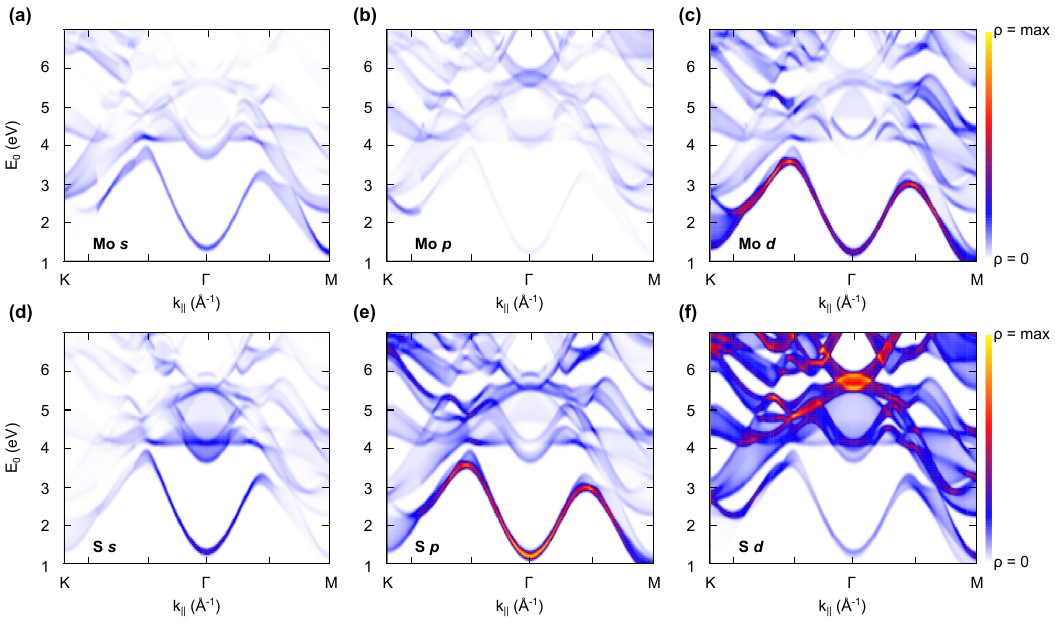}
	\caption{Partial density of states $\rho$ for bulk \mos\ for the lower-energy unoccupied states, shown in dispersion plots ($E_0$ vs $\vec{k}_{//}$ for $K-\Gamma-M$). Upper row: (a) Mo-states of $s$-character ($l=0$). (b)  Mo-states of $p$-character ($l=1$). (c)  Mo-states of $d$-character ($l=2$). Lower row: (d) S-states of $s$-character ($l=0$). (e) S-states of $p$-character ($l=1$). (f) S-states of $d$-character ($l=2$).}
	\label{fig:partialDOS} 
\end{figure*}

Let us next discuss the orbital character of the lower energy bands. Figure \ref{fig:partialDOS} shows the \kp\ and angular-momentum projected partial densities of states of the lower lying bands in semi-infinite \mos\ [see Figs.~\ref{fig:bulk}(a)--\ref{fig:bulk}(c)], for $s$,  $p$, and $d$ character in the muffin-tin spheres of Mo and S. The lowermost band around 0.5\,eV at $\Gamma$ is seen to be predominantly of  Mo-$d$ and S-$p$ character, the latter being the strongest at small \kp. The next states up, around a landing energy of 4 to 5\,eV at $\Gamma$, are of mainly S-$d$ with an admixture of S-$s$ character. This is consistent with our conclusion that these are (unoccupied) interlayer resonances, mainly localized between the lower S-sheet of one \mos\ layer and the upper S-sheet of the next, which means that they are at a significant distance from the Mo-atoms. We note that the intriguing possibility to determine the position and dispersion of S-states with $d$ character is a clear asset of the ARRES technique.\\
As before, our calculations for \mose\ and \mote\ (not shown) point to similar signatures: a dominant mixture of Mo-$d$ and Se/Te-$p$ character for the lowermost state and mostly Se/Te-$d$ character for the interlayer resonances. However, the in-plane dispersion relation of these resonant states does vary over the three TMDs, as already seen in Fig.~\ref{fig:bulk}. Specifically, \mote\ differs from both \mos\ and \mose, the latter two being rather similar at low energies. Note that also the out-of-plane dispersion is stronger for \mote, which is seen from the larger width of the lowermost band in \mote\ in Fig.~\ref{fig:bulk}(i).

\section{Conclusions \& outlook}
We have presented LEEM and ARRES spectra for a series of Mo-based TMDs (\mos, \mose, \mote) of different thicknesses. We have combined these with calculations on unoccupied bands (electron resonances) in these materials. For the bulk case, these are uniquely identified by their energy dispersion relation and characteristic probability density distributions. Both in our experiments and calculations, we find the dispersion relations for the lower-energy resonances to be similar for all three TMDs, although there are small changes upon going from S to Te. For the lowest band (around $E_0=0.5$\,eV at $\Gamma$ for \mos), the electron probability is predominantly found around the Mo-planes. These states have mostly Mo-$d$ and S-$p$ character. The next lowest band (around $E_0=4$ eV at $\Gamma$ for \mos) has a strong S-$d$ character. The dispersion relation for this band stems from the coupling between interlayer resonances in consecutive TMD layers. This interlayer coupling becomes specifically clear in calculations for mono-, double and triple layer TMD systems. For the two-layer case, a single interlayer resonance appears (around 4\,eV for \mos) that is absent for the monolayer. For the three-layer case, this resonance splits up in two. This is a direct result of finite coupling between the two interlayer states, similar to what happens in three-layer graphene and hBN. And as in these materials, the presence of well-defined interlayer resonances in principle allows for local layer counting for thin TMDs. Due to the small interlayer coupling in TMDs, however, the splitting of the resonances is smaller than for graphene and hBN and in our present measurements, we can only distinguish the monolayer from the multilayer systems convincingly. Spectroscopy using LEEM with improved energy resolution is expected to discriminate between few-layer systems~\cite{DeJong2018pssb,Tromp2023}.\\ 
Combining the present work with recent results, we conclude that ARRES is a method capable of discriminating between several key vdW materials, from hBN via graphene to TMDs. It is an open question whether this also holds for more complicated vdW heterostructures and, specifically, if ARRES can provide further insight into the coupling between resonances in different layers. Finally, we emphasize that knowledge of the unoccupied states just above the vacuum level, as obtained by ARRES, is crucial for a correct interpretation of photoemission spectra, including ARPES. The unoccupied resonances studied here play the role of 'final states' in photoemission and secondary electron emission. \\

\begin{acknowledgments}
\textbf{Acknowledgments:}
We thank Marcel Hesselberth and Douwe Scholma for their indispensable technical support, Vera Janssen for providing the exfoliated \mote\ flake, and Peter Neu for discussions.
This work was supported by the Netherlands Organisation for Scientific Research (NWO/OCW) via a VENI grant (680-47-447, J.J.) and the Frontiers of Nanoscience program, as well as by the Spanish Ministry of Science, Innovation and Universities (MCIU Project No.~PID2022-139230NB-I00).
\end{acknowledgments}

\textbf{Additional Note:}
We note that Fig.\ \ref{fig:density}(a) represents the same calculations as those in Fig.\ 2(a) of Ref.~\cite{Neu2023}. Furthermore, the calculated curves in Fig.\ \ref{fig:IVs}(b) and Fig.\ \ref{fig:density}(b) correspond to the $(0,0)$ data in Fig.\ 2(c) of Ref.~\cite{DeJong2018pssb}, with the exception of the bulk case. We present the adapted versions of these data sets here for clarity, to show the reader a more complete comparison between extended theoretical and experimental results for this material series. Finally, the partial density of states calculations for bulk \mose\ and \mote\ (cf. Fig.~\ref{fig:partialDOS}) are available upon reasonable request.

\bibliography{libraryPRBTMD}

\end{document}